# The Jordan-Brouwer theorem for the digital normal n-space $Z^n$.

Alexander V. Evako.
evakoa@mail.ru

**Abstract.**
In this paper we investigate properties of digital spaces which are represented by graphs. We find conditions for digital spaces to be digital n-manifolds and n-spheres. We study properties of partitions of digital spaces and prove a digital analog of the Jordan-Brouwer theorem for the normal digital n-space $Z^n$.

**Key Words**: axiomatic digital topology, graph, normal space, partition, Jordan surface theorem

## 1. Introduction.

A digital approach to geometry and topology plays an important role in analyzing n-dimensional digitized images arising in computer graphics as well as in many areas of science including neuroscience and medical imaging. Concepts and results of the digital approach are used to specify and justify some important low-level image processing algorithms, including algorithms for thinning, boundary extraction, object counting, and contour filling.
To study properties of digital images, it is desirable to construct a convenient digital n-space modeling Euclidean n-space. For 2D and 3D grids, a solution was offered first by A. Rosenfeld [18] and developed in a number of papers, just mention [1-2,13,15-16], but it can be seen from the vast amount of literature that generalization to higher dimensions is not easy.
Even more, as T. Y. Kong indicates in [14], there are problems even for the 2D and 3D grids. For example, different adjacency relations are used to define connectedness on a set of grid points and its complement. Another approach to overcome connectivity paradoxes and some other apparent contradictions is to build a digital space by using basic structures from algebraic topology such as cell complexes or simplicial complexes, see e.g. [7, 12, 17]. Evidently, this approach is not a purely digital one. Considering this, T. Y. Kong [14] formulated the general problem: Construct a simplest possible theory that gives an axiomatic definition of "well-behaved 3D digital spaces" and allows many results of 3D digital topology to be proved simultaneously for all such spaces.
In the present paper, we use axiomatic definitions in order to introduce topology on a finite or countable set of points without involving objects of continuous nature. The material to be presented below begins with definitions and results related to digital objects in section 2. Section 3 covers definitions of digital normal spaces and properties of normal n-manifolds. Normal n-manifolds are digital counterparts of continuous objects [5]. It is shown that for any contractible spaces A and B contained in a normal n-manifold $M$, spaces *M-A* and *M-B* have the same topology. We prove that M is a normal n-sphere if for any contractible space $A$ contained in $M$, the space *M-A* is contractible. In section 4, we define a partition of a given digital object and study properties of this partition. In particular, we prove that any normal *(n-1)*-sphere $S$ contained in a normal n-sphere $M$ is a separating space in $M$. In section 5, we prove the Jordan-Brouwer theorem for a normal digital n-space $Z^n$ investigated in [4]. For *n=2*, it is the Jordan curve theorem.

## 2. Preliminaries.

To make this paper self-contained, we summarize the necessary information from previous papers.
A simple undirected graph is a pair $G=(V,W)$, where $V=\{v_1,v_2,...v_n,...\}$ is a finite or countable set of points, and $W=\{(v_pv_q),....\}\subseteq V\times V$ is a set of edges provided that $(v_pv_q)=(v_qv_p)$ and $(v_pv_p)\notin W$ [8]. Points $v_p$ and $v_q$ are called *adjacent* $(v_p \sim v_q)$ if $(v_pv_q)\in W$.
We use the notations $v_p \in G$ and $(v_pv_q)\in G$ if $v_p \in V$ and $(v_pv_q)\in W$ respectively if no confusion can result. $H=(V_1,W_1)$ is a *subgraph* of a graph $G=(V,W)$ if $V_1\subseteq V$, $W_1\subseteq W$. Since in this paper we use only subgraphs induced by a set of points, we use the word *subgraph* for an induced subgraph. Obviously, H is obtained from $G$ by deleting points. The subgraph $O(v)$ of a graph $G$ containing all points adjacent to a point $v$ (without v) is called *the rim or the neighborhood of v in G.* The subgraph $U(v)= v\cup O(v)$ is called *the ball*

*of v in G.*

Let *G* be a graph and *H⊆G*. The subgraph *O(H)=$\bigcap_{x \in H} O(x)$ is the rim or the mutual adjacency set of H in G* [4,6]. The *local topology* at a point *v* or at an edge *(uv)* is defined by the structure of *O(v)* or *O(uv)* respectively.

The *join G⊕H* of two graphs G and H with disjoint point sets is the graph that contains *G, H* and all edges joining every point in *G* with every point in *H*.

Contractible transformations of graphs seem to play the same role in this approach as a homotopy in algebraic topology [9, 10, 11]. A graph G is called *contractible* if it can be converted to the trivial graph by a sequence of contractible transformations. A point v of a graph G is said to be *simple* if its rim O(v) is a contractible graph. An edge (vu) of a graph G is said to be *simple* if the joint rim O(vu)=O(v)∩O(u) is a contractible graph. Deletions and attachments of simple points and edges are called *contractible transformations*. Graphs G and H are called *homotopy equivalent or homotopic* if one of them can be converted to the other one by a sequence of contractible transformations.

Homotopy is an equivalence relation among graphs. Contractible transformations retain the Euler characteristic and homology groups of a graph [9-10].

Properties of graphs that we will need in this paper were studied in [6, 9-11].

**Proposition 2.1**
- Let G be a contractible graph and H be a graph. Then H⊕G is a contractible graph.
- Let G be a contractible graph with the cardinality |G|>1. Then it has at least two simple points.
- Let H be a contractible subgraph of a contractible graph G. Then G can be transformed into H sequential deleting simple points.
- Let graphs G and H be homotopic. G is connected if and only if H is connected. Any contractible graph is connected.

For any terminology used but not defined here, see Harary [8].

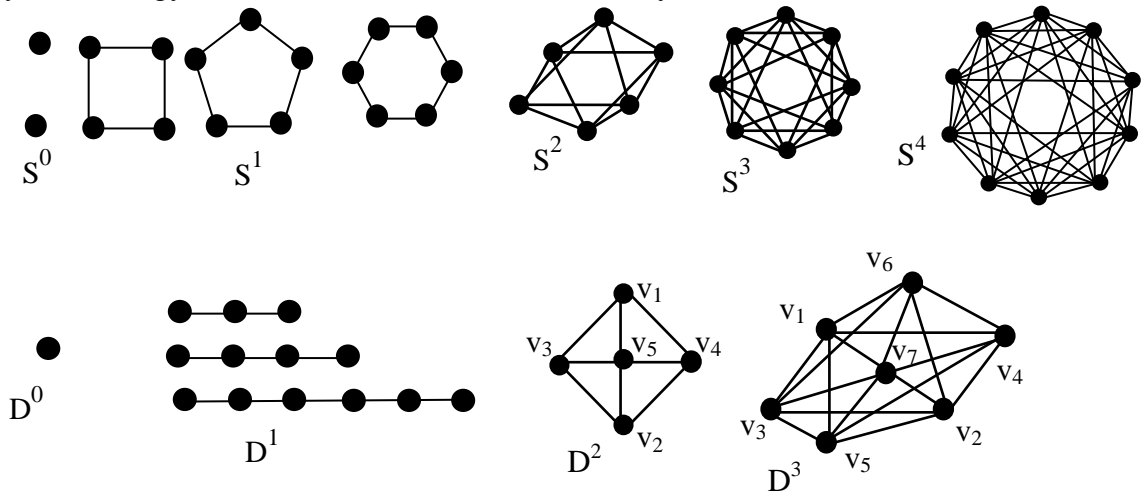

Figure 1. $S^0$, $S^1$, $S^2$, $S^3$ and $S^4$ are normal minimal 0, 1, 2, 3 and 4-spheres. $D^0$, $D^1$, $D^2$ and $D^3$ are normal 0, 1, 2 and 3-disks. For $D^2$, $\partial D^2 = \{v_1, v_2, v_3, v_4\}$, $IntD^2 = v_5$. For $D^3$, $\partial D^3 = \{v_1, v_2, v_3, v_4, v_5, v_6\}$, $IntD^3 = v_7$.

## 3. Basic definitions. Properties of *n*-spheres and *n*-manifolds.

In this approach a *digital space G* is a simple undirected graph *G=(V,W)* whose topological properties in terms of adjacency, connectedness and dimensionality have been studied in [3,4,9-11].

Let $E^n$ be n-dimensional Euclidean space and *x* be a point in it. A neighborhood of *x* is a set *U* which contains an open *n*-ball *B* of center *x*. The boundary of this ball is the *(n-1)*-sphere *S*. Notice that these balls form a basis for the usual topology on $E^n$.

Just as the spheres and balls are used in a definition of dimension in the Euclidean space, the minimal neighborhoods lead to a definition of dimension in graphs.

**Definition 3.1.**
   A *normal 0-dimensional sphere* is a disconnected graph $S^0(a,b)$ with just two points *a* and *b* (fig. 1) [3-4].

To define *n*-spheres and n-manifolds, $n>0$, we will use a recursive definition. Suppose that we have defined normal *k*-spheres for dimensions $1 \leq k \leq n-1$.

**Definition 3.2.**
- A connected space *M* is called *a normal n-dimensional manifold, $n>0$*, if the rim $O(v)$ of any point *v* is a normal *(n-1)*-dimensional sphere.
- A normal n-dimensional manifold *M* is called a *normal n-sphere, $n>0$*, if for any point $v \in M$, the space M-v is contractible (fig. 1)
- Let *M* be a normal *n*-sphere and *v* be a point belonging to *M*. The space $D=M-v$ is called *a normal n-disk* (fig. 1), the space $\partial D = O(v) \subseteq D$ is called the *boundary of D* and the space $IntD = D - \partial D$ is called *the interior of D*.

Figure 1 shows normal n-dimensional sphere, n=1,..4, and n-dimensional disks, n=1,2,3. Normal n-manifolds can be considered as digital images of continuous objects [5].

**Theorem 3.1.**
   Let *M* and *N* be normal *n*- and *m*-spheres. Then $M \oplus N$ is a normal (n+m+1)-sphere.

Proof.
It follows from theorem 5 in [4], that $M \oplus N$ is a normal *(n+m+1)*-space. It is necessary to show that for any point $v \in M \oplus N$, $O(v)$ is a normal *(n+m)*-sphere and $M \oplus N - v$ is a contractible space.
The proof is by induction. For $m+n=0,1$, the theorem is verified directly (fig. 1). Assume that the theorem is valid whenever $m+n \leq k$. Let $m+n=k+1$ and $x \in M$. Denote $O(x)$ the rim of x in $M \oplus N$ and $O(x)_M$ the rim of x in M. Then $O(x) = O(x)_M \oplus N$ is a normal *k*-sphere by the inductive assumption. Consider the space $M \oplus N - x = (M-x) \oplus N$. Since M-x is contractible, then $(M-x) \oplus N$ is contractible according to proposition 2.1 [11]. The proof is complete. □

The following corollary is a direct consequence of theorem 3.1.

**Corollary 3.1.**
- The join $S^n_{min} = S^0_1 \oplus S^0_2 \oplus ... S^0_{n+1}$ of *(n+1)* copies of the normal *0*-dimensional sphere $S^0$ is a normal *n*-sphere (fig. 1).
- A normal n-sphere *S* can be converted into $S^n_{min}$ by contractible transformations.
- $S^n_{min}$ is a normal n-space with the minimal number of points [4].

**Theorem 3.2.**
   Let *M* be a normal *n*-sphere and *A* be a contractible subspace of *M*. Then the space *M-A* is contractible.

Proof.
The proof is by induction on the dimension *n*. For *n=1*, the theorem is verified directly. Assume that the theorem is valid whenever $n<k$. Let $n=k$. Let *M* be a normal *n*-sphere and *A* be a contractible subspace of *M*. Since *A* is contractible, there is a point $x \in A$ such $O(x) \cap A = G$ is contractible (prop. 2.1) [11]. Therefore, *x* is simple in *A*. Since $O(x)$ in *M* is a normal *(n-1)*-sphere, then $O(x) - G = O(x) \cap D$ is also contractible by the induction hypothesis, where $D=M-A$. Hence, $A_1 = A-x$ is a contractible space and $D_1 = D \cup x$ is homotopic to *D*. Acting in the same way, we finally convert the space *A* to a point *v* and the space *D* to *M*-v, where *M*-v and *D* are homotopic to each other, i.e., D is a contractible space. □

**Theorem 3.3.**
   Let *M* be a normal *n*-sphere and *(uv)* be an edge in *M*. Then *M-(uv)* is a normal *n*-disk.

Proof.
Delete from M the point v. The obtained space D=M-v is a normal n-disk with the boundary $\partial D = O(v)$ according to definition 3.2. Let $u \in \partial D$. Glue a point w to D in such a way that the rim $O(w) = \partial D - u$. Since O(w) is a contractible space, then G=D∪w is a contractible space homotopic to D. Hence, G=D∪w is a normal n-disk with the boundary ∂G=u∪w∪O(uw) and the interior IntG=G-∂G by construction. Evidently, D∪w and *M-(uv)* are isomorphic, i.e. M-(uv) is a normal n-disk. The proof is complete. □

Now consider a normal n-manifold, which is a more general construction, then a sphere. The proof of the following theorem is similar to the proof of theorems 3.2 and 3.3 and so is omitted.

**Theorem 3.4.**
Let *M* be a normal *n*-manifold, v be a point in M, (vu) be an edge in M, A and B be contractible subspaces of *M*. Then subspaces M-v, M-(vu) *M-A* and *M-B* are homotopic to each other.

Notice that if a normal *n*-space is not a manifold, then theorem 3.4 does not hold. The following corollary is a direct consequence of theorems 3.2-3.4.

**Corollary 3.2.**
A normal *n*-manifold *M* is a normal *n*-sphere if and only if for any contractible space *A* in *M*, the space *M-A* is contractible.

**Theorem 3.5.**
Let *D=IntD∪∂D* be a normal *n*-disk with the boundary *∂D* and *B* be a subspace of *∂D*. Then the space *D-B* is contractible and any point x belonging to *∂D-B* is simple in *D-A*.
Proof.
Glue a point v to D where O(v)=∂D. Then the obtained space M=v∪D is a normal n-sphere according to definition 3.2. Therefore, the space A=v⊕B⊆M is contractible by proposition 2.1. Hence, M-A=D-B=G is a contractible space by theorem 3.2. For the same reason, if x∈∂D-B then O(x)∩G is a contractible space, i. e., x is a simple point in G. □

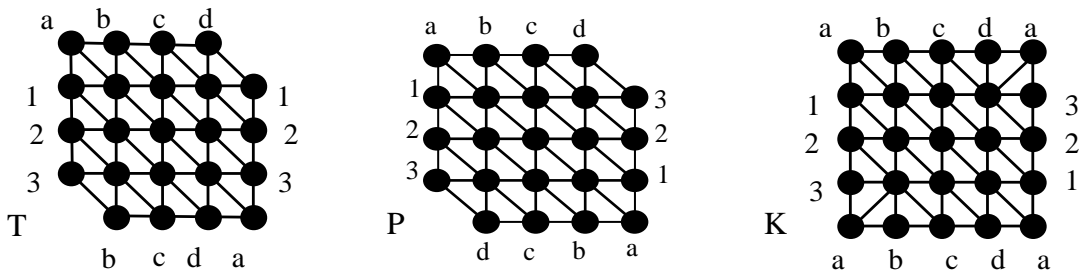

**Figure 2. The normal digital torus T, the normal digital projective plane P and the digital normal Klein bottle K. T, P and K are obtained from a normal 2-disk by identifying boundary points.**

Consider construction of digital normal 2-manifolds by identifying points. In algebraic topology, 2-dimensional manifolds can be built from a fundamental polygon by pairwise identification of its edges. In digital case, a normal 2-disk plays the role of a polygon whose edges are identified. We construct normal digital 2-manifolds by identifying boundary points of a normal 2-disk as it is shown in fig.2, where T is the digital normal torus, P is the digital normal projective plane and K is the digital normal Klein bottle. It is not hard to check that the Euler characteristic and the homology groups of T, P and K are the same as those of their continuous counterparts.
Evidently, there is a normal n-manifold is a special case of a normal n-space. For example, the join $S^0(a,b)⊕T^2$ of the 0-sphere $S^0(a,b)$ and a normal 2-torus $T^2$ is a normal 3-dimensional space, but not a manifold because the rim $O(a)=T^2$ is not a normal 2-sphere. In [4], normal n-spaces were called n-surfaces and properties of such spaces were investigated.
However, it turns out that digital n-manifolds belong to a relatively narrow class of normal digital n-spaces. Most of normal digital spaces, particularly in higher dimensions, are not manifolds in the sense, that the rims of points are not digital (n-1)-dimensional spheres. It will be interesting to find some practical applications for such n-spaces.

## 4. Partition of spaces.

**Definition 4.1.**
> Subspaces *A* and *B* of a connected space *M* are called *disjoint* if any point in *A* is no*n*-adjacent to any point in *B*. We say that the space *C=M-A-*B *separates A and B* and *C is a separating space* in *M*. The union *M=A*∪*C*∪*B* is called a *partition of M*.

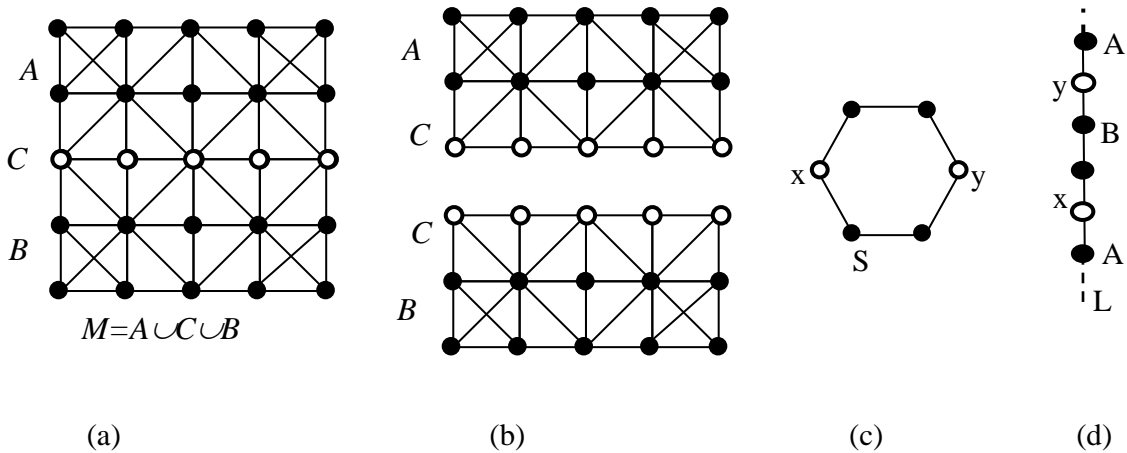

(a)        (b)        (c)        (d)

**Figure 3.** (a) M is a contractible space, the contractible space C is a separating space in M. (b) Spaces A∪C and C∪B are both contractible. (c) A normal 0-sphere $S^0(x,y)$ is a separating space in a normal 1-sphere S. (d) $S^0(x,y)$ is a separating space in the normal 1-line L.

**Theorem 4.1.**
> Let the union *M=A*∪*C*∪*B* be a connected space, *A and B* be no*n*-empty spaces and a contractible space *C* be the separating space in *M*. *M* is contractible if and only if spaces *A*∪*C* and *C*∪*B* are both contractible.

Proof.
Let *M=A*∪*C*∪*B* be a contractible space (fig. 3(a)). Since *C* is contractible, then *M* can be transformed into *C* by sequential deleting simple points [11]. As *A* and *B* are disjoint spaces, then *A*∪*C* and *C*∪*B* are converted into *C* separately. Therefore, *A*∪*C* and *C*∪*B* are both contractible (fig. 3(b)).
For the converse, suppose that subspaces *A*∪*C* and *C*∪*B* are both contractible. This means that *A*∪*C* and *C*∪*B* can be converted into *C* by sequential deleting simple points. Therefore, *A*∪*C*∪*B* can be transformed into *C* by sequential deleting simple points. As *C* is contractible, then *M* is contractible. The proof is completed. □

**Theorem 4.2.**
> Let *M* be a normal *n*-manifold and *S* be a normal (*n-1*)-sphere in *M*. If there is a contractible space *G* in *M* containing *S*, then *S* is a separating space in *M*, *M=A*∪*S*∪*B,* and the space *S*∪*B* (or/and *A*∪*S*) is a normal *n*-disk.

Proof.
The proof is by induction on the dimension *n*. For n=1, the theorem is plainly true. Assume that the theorem is valid whenever *n≤k*. Let *n=k+1*.
According to proposition 2.1, a contractible space with more than one point has at least two simple points. Therefore, there are at least two simple points in *G*. Sequentially delete from *G* simple points belonging to *G-S*. In the obtained space *D*⊆*M* , take a point *x*∈*S* (fig. 4(a)). Since *S-x* is a contractible space by definition 3.2, then D can be converted to *S-x* by sequential deleting simple points. Therefore, *x* is necessarily a simple point in *D*, i.e., *O(x)*∩*D* is contractible. Evidently, $S_x$=O(x)∩S⊆O(x)∩D  is a normal *(n-2)*-sphere (definition 3.2). As $x \in M$, then $O(x) \cap M$ is a normal $(n-1)$-sphere. Hence $O(x)=A_x \cup S_x \cup B_x$ is a partition of $O(x)$, $S_x$ is a separating space in $O(x)$, and $A_x \cup S_x$ and $S_x \cup B_x$ are normal $(n-1)$-disks by the induction hypothesis. Evidently, $O(x) \cap D= E_x=S_x \cup B_x$.

Build the space $D_1=(D-x) \cup y$ by gluing a point $y$ to $D$ in such a way that $O(y)=S-x$ and deleting $x$. By construction, $D_1$ is a contractible space and $S_1=E_x \cup y$ is a normal $(n-1)$-sphere (fig. 4(b)). Let $z \in B_x$. For the same reason as above, $O(z) \cap D_1$ is a normal $(n-1)$-disk. Therefore, $O(z) \cap D=x \cup (O(z) \cap D_1)$ is a normal $(n-1)$-sphere by construction (fig. 4(c)). Since $O(z) \cap M$ is a normal $(n-1)$-sphere and $O(z) \cap D \subseteq O(z) \cap M$, then $O(z) \cap M = O(z) \cap D \subseteq D$ [5]. For the same reason as above for any point $v \in D-S$, $O(v) \subseteq D$. Hence, D is a normal n-disk in accordance with definition 3.2 The proof is complete. □

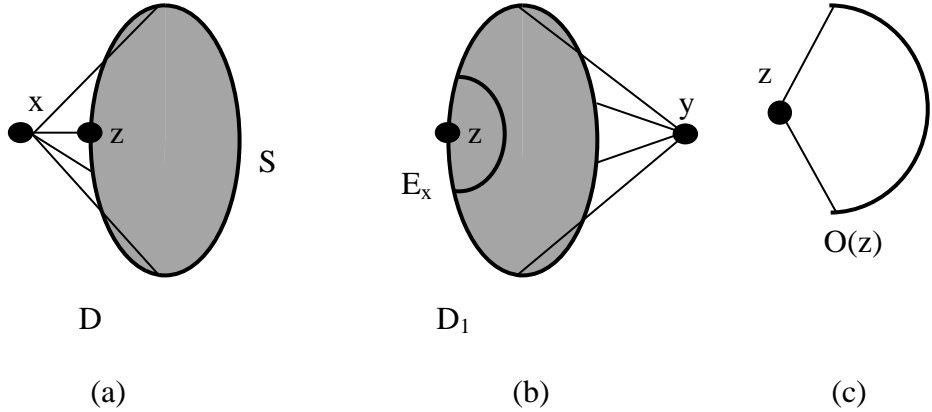

Figure 4. (a) *D is a contractible space containing S. $O(x) \cap D$ is a normal $(n-1)$-disk.* (b) *The space $D_1=(D-x) \cup y$ is contractible, $O(z) \cap D_1$ is a normal $(n-1)$-disk.* (c) *$O(z)= O(z) \cap D$ is a normal $(n-1)$-sphere.*

**Theorem 4.3.**
  Let *M* be a normal *n*-sphere and *S* be a normal $(n-1)$-sphere in *M*. Then *S* is a separating space in $M=A \cup S \cup B$, and spaces $A \cup S$ and $S \cup B$ are both normal $(n-1)$-disks (fig. 3(c)).

Proof.

Let a point $x \in M-S$. By definition 3.2, $M-x$ is a contractible space containing *S*. Then $S \cup B$ is a normal *n*-disk lying in $M-x$, $M=A \cup S \cup B$ is a partition of *M* and *S* is a separating space in *M* according to theorem 4.2. Let a point $y \in B$. For the same reason as above, $A \cup S$ is a normal *n*-disk lying in $M-y$, $M=A \cup S \cup B$ is a partition of *M* and *S* is a separating space in *M*. Thus, *S* is a separating space in $M=A \cup S \cup B$, and spaces $A \cup S$ and $S \cup B$ are a normal *n*-disks. □

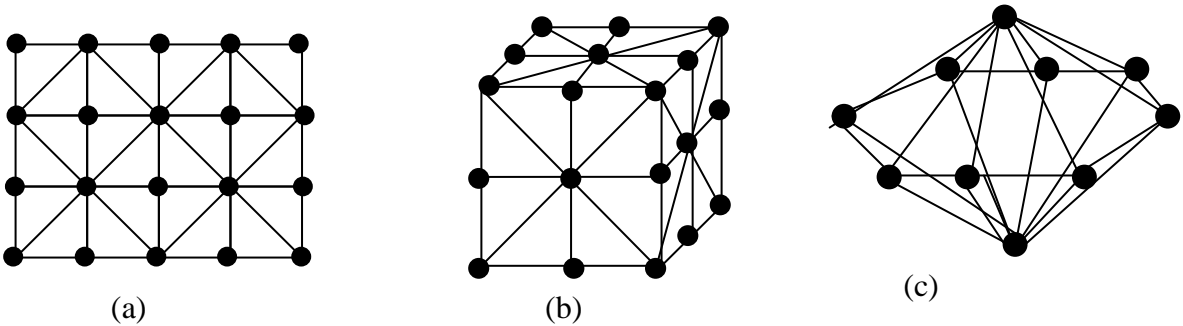

Figure 5. (a) The normal 2-space $Z^2$. (b) The rim of a pure point in $Z^3$. (c) The rim of a mixed point in $Z^3$.

**Theorem 4.4.**

Let $G=A\cup S$ and $H=C\cup B$ be normal $n$-disks and their boundaries $\partial G=S$ and $\partial H=C$ be isomorphic. Then the space $M=A\cup S\cup B$ obtained by identifying points in $S$ with corresponding points in $C$ is a normal $n$-sphere and $S$ is a separating space in $M$.

Proof.
The proof is by induction on the dimension $n$. For $n=1$, the theorem is verified directly. Assume that the theorem is valid whenever $n<k$. Let $n=k$ and $M=A\cup S\cup B$. Take a point $x\in S\subseteq M$. Then O(x) is a normal (n-1)-sphere in M by the induction hypothesis. Since for any $x\in S$, O(x) is a normal (n-1)-sphere in M, then M is a normal n-manifold by definition 3.2. Let $y\in S\subseteq M$ and consider the space M-x= $A\cup(S-x)\cup B$. Evidently, S-x is a separating space in M-x and spaces $A\cup(S-x)$, $(S-x)\cup B$ and S-x are contractible spaces by construction. Therefore, M-x is a contractible space by theorem 4.1. Hence, M is a normal n-sphere by definition 3.2. □

## 5. The Jordan surface theorem for the normal n-space $Z^n$.

The next definition concerns the normal space $Z^n$ investigated in [4].

**Definition 5.1.**
- Let $Z^n$ be a set of points with integer coordinates in n-dimensional Euclidean space $E^n$. We say that *$Z^n$ is the normal n-space* if points $x=(x_1,..,x_n)$ and $y=(y_1,..,y_n)$ are adjacent in $Z^n$ if $|x_i-y_i|\leq 1$ and $t(x)_i \leq t(y)_i$ for $i=1,\ldots,n$, where $t(x)_i=x_i \bmod 2$, $t(y)_i=y_i \bmod 2$.
- We say that a subspace $U=\{x=(x_1,..,x_n)|\ a_i\leq x_i\leq b_i,\ a_i<b_i,\ i=1,\ldots n\}$ is an n-box in $Z^n$..

$Z^n$ is considered here as a digital counterpart of $E^n$.
A point $x\in Z^n$ is called pure if coordinates of x are all odd or all even. All other points are called mixed. It was shown in [4] that $Z^n$ is the topological product of n copies of the set Z of integers with the Khalimsky topology. Note that the notion of the adjacency set of a point given in [4] essentially generalizes the notion used in [15] for Alexandroff spaces. The following property was proved in [4].

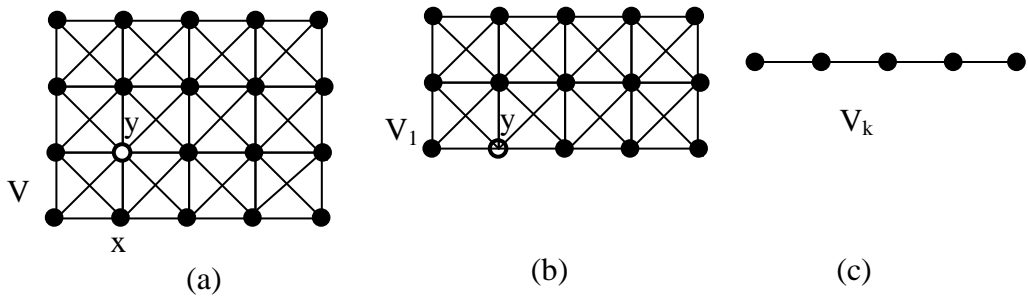

Figure 6. (a) The 2-box V in the complete 2-space $Y^2$. (b) $V_1$ is obtained from V by deleting points for which $x_2=a_2$. (c) $V_k$ is a 1-box in $Y^1$.

**Theorem 5.1.**
The rim O(x) of any point $x\in Z^n$ is a normal (n-1)-sphere, i.e. $Z^n$ is a normal n-manifold.

Figure 5(a) shows a part of the digital plane $Z^2$ which is the Alexandroff space, figure 5(b) depicts the rim of a pure point in $Z^3$, figure 5(c) shows the rim of a mixed point in $Z^3$. It is easy to see that $Z^n$ contains hyperplanes of dimension $m<n$. For example, the space G contained in $Z^n$ and defined by equations $x_i=a_i$, $i=1,\ldots k$, $k<n$, is the normal (n-k)-space $Z^{n-k}$.

**Definition 5.2.**
- We say that a set $Y^n$ of points with integer coordinates in $E^n$ is the complete n-space if points $x=(x_1,..,x_n)$ and $y=(y_1,..,y_n)$ are adjacent if $|x_i-y_i|\leq 1$, $i=1,\ldots,n$.
- We say that a subspace $V=\{x=(x_1,..,x_n)|\ a_i\leq x_i\leq b_i,\ a_i<x_i<b_i,\ i=1,\ldots n\}$ is an n-box in $Y^n$.

Fig. 6(a) shows a 2-box V in $Y^2$.

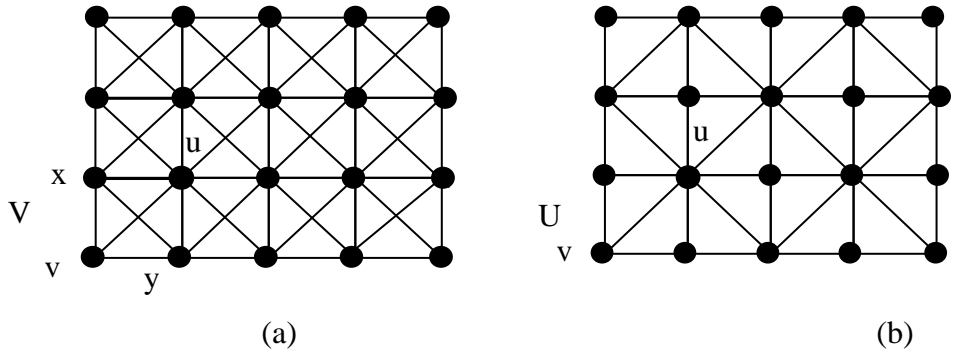

**Figure 7.** (a) The 2-box V in $Y^2$. (b) The 2-box U in $Z^2$ obtained from V by deleting simple edges.

**Proposition 5.1.**
 An n-box $V=\{x=(x_1,..,x_n)| a_i \le x_i \le b_i, i=1,...n\}$ in $Y^n$ is a contractible space.

Proof.
The proof is by induction on the dimension n. For n=1, the statement is verified directly. Assume that the statement is valid whenever $n \le k$. Let $n=k+1$. Take a point $x=(x_1,x_2,..,x_{n-1},x_n=a_n)$ belonging to V. By construction, $O(x) \cap V$ contains the point $y=(x_1,x_2,..,x_{n-1},x_n=a_n+1)$ (fig. 6(a)), which is adjacent to all other points belonging to $O(x) \cap V$, i.e. $O(x) \cap V = y \oplus G$. Therefore, $O(x) \cap V$ is a contractible space, i.e. x is a simple point and is deleted from V according to proposition 2.1. Let $V_1$ is the space obtained by deleting all such point (fig. 6(b)). For the same reason as above, all points for which $x_n=a_n+1$ can be deleted from $V_1$. Finally, we obtain the (n-1)-box $V_k=\{x_1,..,x_{n-1},x_n=b_n| a_i \le x_i \le b_i, i=1,...n-1\}$ (fig. 6(c)), which is a contractible space by the induction hypothesis. This completes the proof. □

**Proposition 5.2.**
 An n-box $U=\{x=(x_1,..,x_n)| a_i \le x_i \le b_i, a_i < b_i, i=1,...n\}$ in $Z^n$ is a contractible space.

Proof.

Let $V=\{x=(x_1,..,x_n)| a_i \le x_i \le b_i, i=1,...n\}$ be n-box in $Y^n$ with the same set of points as U of $Z^n$ (fig. 7(a), fig. 7(b)). With no loss of generality, assume that points x and y belong to V (and U), $x=(\{1\}_k,\{0\}_m,\{e\}_p)$ and $y=(\{0\}_k,\{1\}_m,\{e\}_p)$, k>0, m>0, p>0, k+m+p=n. By construction, x and y are adjacent in V and are not adjacent in U.
On the other hand, points $v=(\{1\}_k,\{1\}_m,\{e\}_p)$ and $u=(\{0\}_k,\{0\}_m,\{e\}_p)$ belong to V (or at least on of them), are adjacent to x, y and any point z belonging to $O(xy) \cap V$, i.e., $O(xy) \cap V = v \oplus u \oplus A$ (fig. 7(a)) is a contractible space according to proposition 2.1. Hence, the edge (xy) is simple in V and can be deleted from V. After deleting all such edges, we obtain the space U. Since V is converted to U by sequential simple edges, then V and U are homotopy equivalent, i.e., u is a contractible space. □

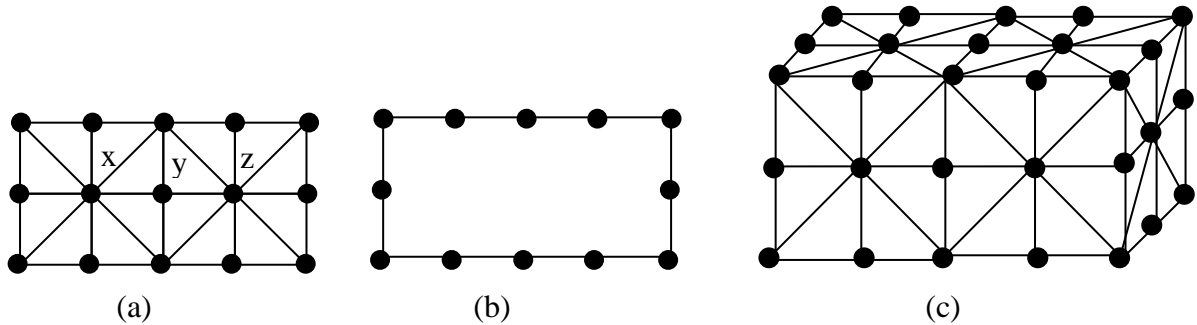

**Figure 8.** (a) A normal *2*-disk $D=O(x) \cup O(y) \cup O(z)$ contained in $Z^2$. (b) A normal 1-sphere $S=D-x-y-z$. (c) A normal 2-sphere contained in $Z^3$.

$Z^n$ contains normal (n-1)-spheres different from the rims of points. It is not hard to check directly that the boundary of an n-box U={x| $2a_i \le x_i \le 2b_i$, i=1,…n, $a_i<b_i$}⊆$Z^n$ is a normal (n-1)-sphere. As an example, take points x=(0,0,…0), y=(1,0,…0) and z=(2,0,…0). Then for $Z^2$, a normal 1-disk D=O(x)∪O(y)∪O(z) is depicted in fig. 8(a), a normal 1-sphere S=[O(x)∪O(y)∪O(z)]-x-y-z, which is the boundary of D, is depicted in fig. 8(b). For $Z^3$, a normal 2-sphere is shown in fig. 8(c).

Now we will prove the Jordan-Brouwer theorem for the normal n-space $Z^n$.

**Theorem 5.1.**
  Let S be a normal (n-1)-sphere in $Z^n$. Then S is a separating space in $Z^n$, $Z^n$=A∪S∪B is a partition and the space S∪B is a normal n-disk with the boundary S.

Proof.

$Z^n$ is a normal n-manifold as it was shown in [4]. Let S⊆$Z^n$ be a normal (n-1)-sphere in $Z^n$. Since S consists of a finite number of points, then there is an n-box U⊆$Z^n$, such that S⊆U. According to proposition 5.2, U is a contractible space. In accordance with theorem 4.2, S is a separating space in $Z^n$, $Z^n$=A∪S∪B is a partition and the space S∪B is a normal n-disk with the boundary S. The proof is complete. □

Fig.9(a) shows a 2-box U⊆$Z^2$ containing the 1-sphere S, which is a separating space in $Z^2$.

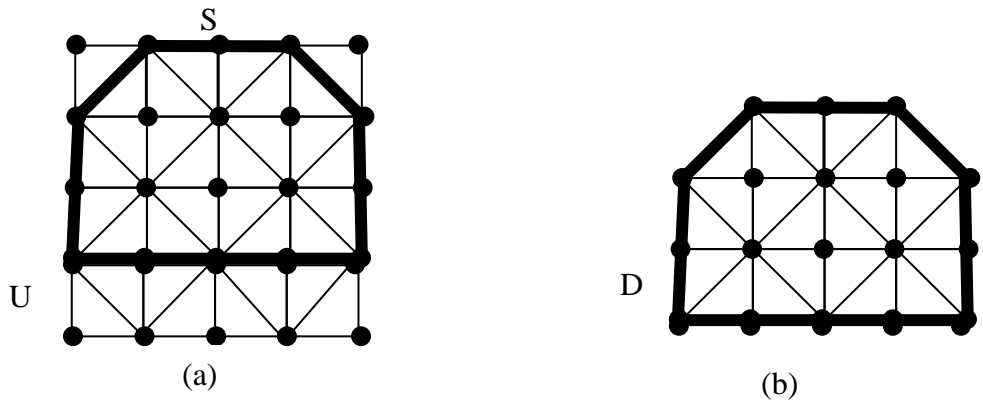

(a)  (b)

**Figure 9. (a) U is a 2-box containing S. (b) The space D obtained from U by deleting simple points from U-S. $Z^2$=A∪S∪B, where A=$Z^2$-D, B=D-S.**

## 6. Conclusion.

Using simple axiomatic definitions we introduce digital topology on a finite or countable set of points. We build normal digital n-dimensional manifolds and investigate their properties. As an example, by identifying points we construct normal digital 2-dimensional spaces: the Klein bottle, the torus and the projective plane. We define the partition of a digital space and prove that if a normal digital (n-1)-sphere S is contained in a normal digital n-manifold M and there is a contractible space G such that S⊆G⊆M, then S is a separating space in M. Finally, we prove the Jordan-Brouwer theorem for the normal digital n-space $Z^n$ studied in [4].

## References.